\documentclass[%
 reprint,
 amsmath,amssymb,
 aps,
prb,
]{revtex4-1}

\usepackage{graphicx}
\usepackage{dcolumn}
\usepackage{bm}
\usepackage{hyperref}
\usepackage[mathlines]{lineno}

\begin{document}

\preprint{APS/123-QED}

\title{Defect-induced large spin-orbit splitting in the monolayer of PtSe$_2$}

\author{Moh. Adhib Ulil Absor}
\email{adib@ugm.ac.id} 
\affiliation{Department of Physics, Universitas Gadjah Mada BLS 21 Yogyakarta Indonesia.}%

\author{Iman Santosa}
\affiliation{Department of Physics, Universitas Gadjah Mada BLS 21 Yogyakarta Indonesia.}%

\author{Harsojo}
\affiliation{Department of Physics, Universitas Gadjah Mada BLS 21 Yogyakarta Indonesia.}%

\author{Kamsul Abraha}
\affiliation{Department of Physics, Universitas Gadjah Mada BLS 21 Yogyakarta Indonesia.}%

\author{Fumiyuki Ishii}%
\affiliation{Faculty of Mathematics and Physics Institute of Science and Engineering Kanazawa University 920-1192 Kanazawa Japan.}%

\author{Mineo Saito}
\affiliation{Faculty of Mathematics and Physics Institute of Science and Engineering Kanazawa University 920-1192 Kanazawa Japan.}%

\date{\today}

\begin{abstract}
The effect of spin-orbit coupling (SOC) on the electronic properties of monolayer (ML) PtSe$_2$ is dictated by the presence of the crystal inversion symmetry to exhibit spin polarized band without characteristic of spin splitting. Through fully-relativistic density-functional theory calculations, we show that large spin-orbit splitting can be induced by introducing point defects. We calculate stability of native point defects such as a Se vacancy (V$_{\texttt{Se}}$), a Se interstitial (Se$_{i}$), a Pt vacancy (V$_{\texttt{Pt}}$), and a Pt interstitial (Pt$_{i}$), and find that both the V$_{\texttt{Se}}$ and Se$_{i}$ have the lowest formation energy. We also find that in contrast to the Se$_{i}$ case exhibiting spin degeneracy in the defect states, the large spin-orbit splitting up to 152 meV is observed in the defect states of the V$_{\texttt{Se}}$. Our analyses of orbital contributions to the defect states show that the large spin splitting is originated from the strong hybridization between Pt-$d_{x{^2}+y{^2}}+d_{xy}$ and Se-$p_{x}+p_{y}$ orbitals. Our study clarifies that the defects play an important role in the spin splitting properties of the PtSe$_2$ ML, which is important for designing future spintronic devices.    
\end{abstract}

\pacs{Valid PACS appear here}
\keywords{Suggested keywords}
\maketitle

\section{INTRODUCTION}
Much of the recent interest in spintronics has been focused on the manipulation of non-equilibrium materials using spin-orbit coupling (SOC) \cite {Wolf,Zutic}. When the SOC occures in system with sufficiently low crystalline symmetry, an effective magnetic field $\textbf{B}_{\texttt{eff}}\propto [\nabla V(\textbf{r})\times \textbf{p}]$ is induced \cite {Rashba,Dresselhauss}, where $V(\textbf{r})$ denotes the crystal potential and $\textbf{p}$ is the momentum, that leads to spin splitting even in nonmagnetic materials. Current-induced spin polarization \cite {Kuhlen} and spin Hall effect \cite {Qi} are important examples of spintronics phenomena where the SOC plays an important role. For spintronic devices operation \cite {Datta}, semiconductor materials having large spin splitting are highly desirable \cite {Ishizaka}. Besides their electronic manipulability under gate voltages \cite {Lu,Nitta}, semiconductors with the large spin splitting enable us to allow operation as a spintronic devices at room temperature \cite {Yaji,King}.  

Two dimensional (2D) transition metal dichalcogenideas (TMDs) family is promising candidates for spintronics due to the strong SOC \cite {Kosminder,Zhu,Latzke,Liu_Bin, Absor4}. Most of the 2D TMDs family have a graphene-like hexagonal crystal structures consisting of transition metal atoms ($M$) sandwiched between layers of chalcogen atoms ($X$) with $MX_{2}$ stoichiometry. However, depending on the chalcogen stacking, there are two stable forms of the $MX_{2}$ in the ground state, namely a $H$ phase having trigonal prismatic hole for metal atoms, and a $T$ phase that consists of staggered chalcogen layers forming octahedral hole for metal atoms \cite {Cudazzo}. In the $H-MX_{2}$ monolyer (ML) systems such as molybdenum and tungsten dichalcogenides (MoS$_{2}$, MoSe$_{2}$, WS$_{2}$, and WSe$_{2}$), the absence of inversion symmetry in the crystal structure together with strong SOC in the 5$d$ orbitals of transition metal atoms leads to the fact that a large spin splitting has been established \cite {Kosminder,Zhu,Latzke,Liu_Bin,Absor4}. This large spin splitting is believed to be responsible for inducing some of interesting phenomena such as spin Hall effect \cite{Cazalilla,Ma}, spin-dependent selection rule for optical transitions \cite{Chu}, and magneto-electric effect in TMDs \cite {Gong}. Furthermore, the long-lived spin relaxation and spin coherence of electrons have also been reported on various $H-MX_{2}$ TMDs ML such as MoS$_{2}$ ML \cite {Schmidt,L_Yang} and WS$_{2}$ ML \cite {L_Yang}, that could be implemented as energy-saving spintronic devices.

Recently, PtSe$_{2}$ ML,  a 2D TMDs ML with $T-MX_{2}$ ML structures, has attracted much attention since it was successfully synthesized by a single step fabrication methods, a direct selenization at the Pt(111) substrate \cite {LWang}, which is in contrast to conventional fabrication methods used in the various $H-MX_{2}$ TMDs ML such as exfoliation \cite {Radisavljevic} or chemical vapor deposition (CVD) \cite {Vander,Elias}. Moreover, the high electron mobility up to 3000 cm$^{2}$/V/s has been experimentally observed on the PtSe$_{2}$ ML, which is the largest among the studied TMDs ML \cite {XZhang}, and thus is of great interest for electronic applications. However, the PtSe$_{2}$ ML has the crystal inversion symmetry, and consequently, the SOC leads to spin polarized bands without characteristic of the spin splitting. This is supported by the fact that the absence of the spin splitting has been experimentally observed by Yao $et$. $al$. using spin- and angle-resolved photoemission spectroscopy (spin-ARPES) \cite {Yao}. Because the absence of the spin splitting in the PtSe$_{2}$ ML possess natural limit for spintronic applications, it is highly desirable to find a method to generate the spin splitting in the PtSe$_{2}$ ML, which is expected to enhance its functionality for spintronics. 

\begin{figure*}
	\centering
		\includegraphics[width=0.7\textwidth]{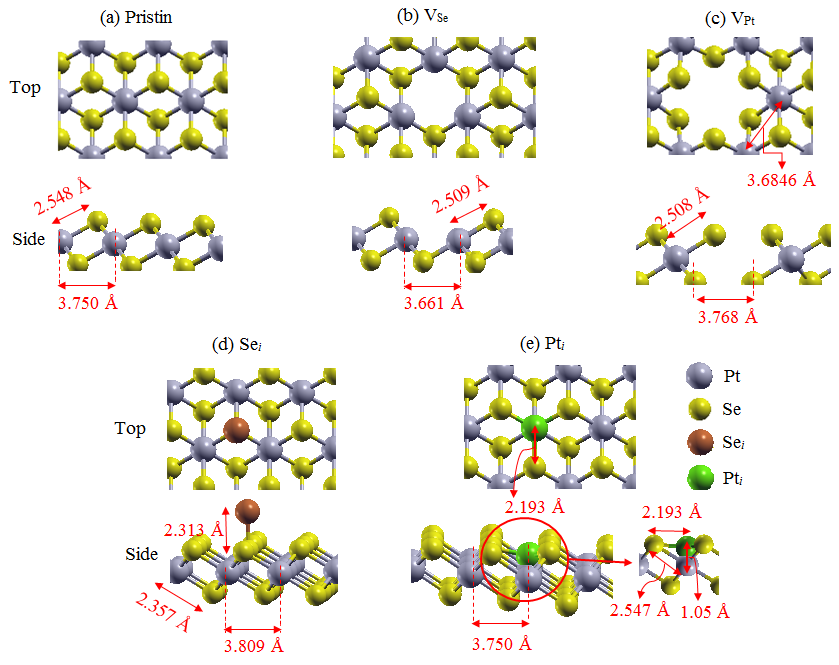}
	\caption{The relaxed structures of native point defects induced by vacancy and interstitial in the PtSe$_2$ ML compared with the pristin system: (a) a pristin, (b) a Se vacancy (V$_{\texttt{Se}}$), (c) a Se interstitial (Se$_{i}$), (d) a Pt vacancy (V$_{\texttt{Pt}}$), and (e) a Pt interstitial (Pt$_{i}$). The Pt-Se, Pt-Pt, Se$_{i}$-Se, and Pt$_{i}$-Pt bond lengths are indicated by the red arrows.}
	\label{figure:Figure1}
\end{figure*}

In this paper, by using fully-relativistic density-functional theory (DFT) calculations, we show that large spin-orbit splitting in the PtSe$_{2}$ ML can be induced by introducing point defects. We calculate stability of native point defects such as a Se vacancy (V$_{\texttt{Se}}$), a Se interstitial (Se$_{i}$), a Pt vacancy (V$_{\texttt{Pt}}$), and a Pt interstitial (Pt$_{i}$), and find that both the V$_{\texttt{Se}}$ and Se$_{i}$ have the lowest formation energy. By taking into account the effect of the SOC in our DFT calculations, we find that in contrast to the Se$_{i}$ case having spin degeneracy in the defect states, the large spin-orbit splitting up to 152 meV is observed on the defect states of the V$_{\texttt{Se}}$. We clarify the origin of the spin splitting by considering orbital contributions to the defect states, and find that the large spin splitting is mainly originated from the strong hybridization between Pt-$d_{x{^2}+y{^2}}+d_{xy}$ and Se-$p_{x}+p_{y}$ orbitals. Finally, a possible application of the present system for spintronics will be discussed.

\section{Computational Details}

We performed first-principles electronic structure calculations based on the density functional theory (DFT) within the generalized gradient approximation (GGA) \cite {Perdew} using the OpenMX code \cite{Openmx}. We used norm-conserving pseudopotentials \cite {Troullier}, and the wave functions are expanded by the linear combination of multiple pseudoatomic orbitals (LCPAOs) generated using a confinement scheme \cite{Ozaki,Ozakikino}. The orbitals are specified by Pt7.0-$s^{2}p^{2}d^{2}$ and Se9.0-$s^{2}p^{2}d^{1}$, which means that the cutoff radii are 7.0 and 9.0 bohr for the Pt and Se atoms, respectively, in the confinement scheme \cite{Ozaki,Ozakikino}. For the Pt atom, two primitive orbitals expand the $s$, $p$, and $d$ orbitals, while, for the Se atom, two primitive orbitals expand the $s$ and $p$ orbitals, and one primitive orbital expands $d$ orbital. Spin-orbit coupling (SOC) was included in our DFT calculations.

Bulk PtSe$_{2}$ crystallizes in a centrosymmetric crystal associated with a $T$ structure ($T-MX_{2}$), having space group $P\overline{3}mI$ space group for the global structure and polar group $C_{3v}$ and $D_{3d}$ for the Se and Pt sites, respectively. In the monolayer (ML) phase, one Pt atom is sandwiched between two Se atoms forming an octahedral hole for transition metal atoms and shows trigonal structure when projected to the (001) plane [Fig. 1(a)]. In our DFT calculations, we used a periodic slab to model the PtSe$_2$ ML, where a sufficiently large vacuum layer (20 \AA) is used to avoid interaction between adjacent layers. The geometries were fully relaxed until the force acting on each atom was less than 1 meV/\AA. We find that the calculated lattice constant of the PtSe$_{2}$ ML is 3.75 \AA, which is in good agreement with the experiment (3.73 \AA \cite {LWang}) and previous theoretical calculations (3.75 \AA \cite {LZhuang,WZhang,Zulfiqar}).

We then introduces native point defects consisting of a Se vacancy (V$_{\texttt{Se}}$), a Se interstitial (Se$_{i}$), a Pt vacancy (V$_{\texttt{Pt}}$), and a Pt interstitial (Pt$_{i}$) [Figs. 1(b)-(e)]. To model these point defects, we construct a 4x4x1 supercell of the pristin the PtSe$_{2}$ ML with 48 atoms. The larger supercell (5x5x1 and 6x6x1 supercells) is used to test our calculational results, and we confirmed that it does not affect to the main conclusion. We calculate formation energy to confirmed stability of these point defects by using the following formula \cite {Freysoldt}:
\begin{equation}
\label{1}
E_{f}=E_{\texttt{defect}}-E_{\texttt{Perfect}}+\sum_{i}n_{i}\mu_{i}.
\end{equation}
In Eq. (1), $E_{\texttt{defect}}$ is the total energy of the defective system, $E_{\texttt{Perfect}}$ is the total energy of the perfect system, $n_{i}$ is the number of atom being added or removed from the perfect system, and $\mu_{i}$ is the chemical potential of the added or removed atoms corresponding to the chemical environment surrounding the system. Here, $\mu_{i}$ obtains the following requirements:
\begin{equation}
\label{2a}
E_{PtSe_{2}}-2E_{Se}\leq \mu_{Pt}\leq E_{Pt},
\end{equation}
\begin{equation}
\label{2b}
\frac{1}{2}(E_{PtSe_{2}}-E_{Pt})\leq \mu_{Se}\leq E_{Se}.
\end{equation}
Under Se-rich condition, $\mu_{Se}$ is the energy of the Se atom in the bulk phase (hexagonal Se, $\mu_{Se}=\frac{1}{3}E_{Se-hex}$) which corresponds to the lower limit on Pt, $\mu_{Pt}=E_{PtSe_{2}}-2E_{Se}$. On the other hand, in the case of the Pt-rich condition, $\mu_{Pt}$ is ubject to the energy of the Pt atom in the bulk phase (fcc Pt, $\mu_{Pt}=\frac{1}{4}E_{Pt-fcc}$) corresponding to the lower limit on Se, $\mu_{Pt}=\frac{1}{2}(E_{PtSe_{2}}-E_{Pt})$.

\section{RESULT AND DISCUSSION}

First, we examine energetic stability and structural relaxation in the defective PtSe$_{2}$ ML systems. Tabel 1 shows the calculated results of the formation energy for the point defects (V$_{\texttt{Se}}$, V$_{\texttt{Pt}}$, Se$_{i}$, Pt$_{i}$) corresponding to the Pt-rich and Se-rich conditions. Consistent with previous studies \cite {Komsa1,WZhang,Zulfiqar}, we find that the V$_{Se}$ and Se$_{i}$ have the lowest formation energy in both the Pt-rich and Se-rich conditions, indicating that both systems are the most stable point defects formed in the PtSe$_{2}$ ML. The found stability of the V$_{\texttt{Se}}$ and Se$_{i}$ is consistent with previous reports that the chalcogen vacancy and interstitial can be easily formed in the TMDs ML as found in the MoS$_{2}$ \cite {Noh,Komsa1,Komsa2}, WS$_{2}$ \cite {Li}, and ReS$_{2}$ \cite {Horzum}. In contrast, the formation of the other point defects (V$_{\texttt{Pt}}$ and Pt$_{i}$) is highly unfavorable due to the required electron energy. Because the Pt atom is covalently bonded to the six neighboring Se atoms, adding or removing the Pt atom is stabilized by destroying the Se sublattice, thus increases the formation energy. 

\begin{table}[h!]
\caption{Formation energy (in eV) of various point defects in the PtSe$_{2}$ ML corresponding to the Se-rich and Pt-rich conditions. The theoretical data from the previous report are given for a comparison.} 
\centering 
\begin{tabular}{c c c c} 
\hline\hline 
Point defects & Pt-rich (eV) & Se-rich (eV)  & Reference \\ 
\hline 
V$_{\texttt{Se}}$ & 1.27 & 1.84 & This work \\ 
         & 1.24 &   1.83   &   Ref. \cite{WZhang} \\
         & - & 1.82 & Ref. \cite{Komsa1} \\				
V$_{\texttt{Pt}}$	& 3.06 & 4.28 & This work \\
				 & 3.00 & 4.19 & Ref. \cite{WZhang} \\
         & - & 3.7 & Ref. \cite{Zulfiqar} \\ 
Se$_{i}$& 2.01 & 1.98 &  This work \\ 
Pt$_{i}$& 4.68 & 3.45 & This work\\ 
\hline\hline 
\end{tabular}
\label{table:Table 2} 
\end{table}

Due to the relaxation, position of the atoms around the point defects marginally changes from the position of the pristin atomic positions. In the case of the V$_{\texttt{Se}}$, one Se atom in a PtSe$_{2}$ ML is removed in a supercell [Fig. 1(b)], and consequently, three Pt atoms surrounding the vacancy is found to be relaxed moving close to each other. Around the V$_{\texttt{Se}}$ site, the Pt-Se bond length at each hexagonal sides has the same value of about 2.509 \AA. As a result, trigonal symmetry suppresses the V$_{\texttt{Se}}$ to exhibit the $C_{3v}$ point group [Fig. 1(b)]. Similar to the V$_{\texttt{Se}}$ case, the V$_{\texttt{Pt}}$ retains threefold rotation symmetry [Fig. 1(c)], yielding the $D_{3h}$ point group. We find that the Pt-Se bond length in the V$_{\texttt{Pt}}$ is 2.508 \AA, which is slightly lower than that of the pristin system (2.548 \AA).

The geometry of the Se$_{i}$ undergoes significant distortion from the pristin crystal, but the symmetry itself remains unchanged [Fig. 1(d)]. Here, we investigate various atomic configurations for the Se$_{i}$, and find that the Se adatom structure on top of a host Se atom is the most stable configuration. In this configuration, we find that the Se-Se$_{i}$ bond length is 2.313 \AA. We also find two other meta-stable configurations of the Se$_{i}$: (i) bridge position of the Se$_{i}$ with two host Se atoms on the surface (2.94 eV higher in energy) and (ii) hexagonal interstitial in the Pt layer bonding with the three host Pt atoms (5.95 eV higher in energy). The Pt$_{i}$ is the most complicated case among the choosen point defects. There are five atomic configurations of the Pt$_{i}$, and we confirmed that the Pt-Pt$_{i}$ split interstitial along the $c$ direction [Fig. 1(e)] is the most stable configuration. Four other meta-stable configurations are (i) the bridge configuration of Pt$_{i}$ in between two surface Se atoms (0.95 eV higher in energy), (ii) the Pt$_{i}$ at hexagonal hollow center on the surface (1.44 eV higher in energy), (iii) the Pt$_{i}$ hexagonal hollow center in the Pt layer (3.05 eV higher in energy), and (iv) the Pt$_{i}$ on top of a surface Se atom (3.15 eV higher in energy). In the Pt-Pt$_{i}$ split interstitial configuration, we find that the Pt-Pt$_{i}$ bond length is 1.05 \AA.

\begin{figure*}
	\centering
		\includegraphics[width=0.75\textwidth]{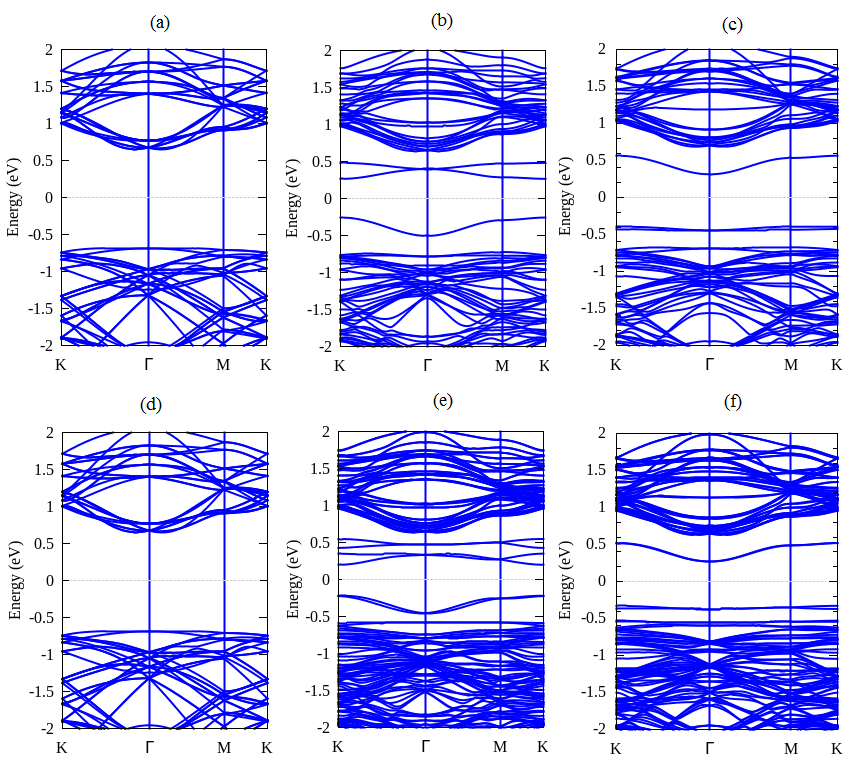}
	\caption{The electronic band structure of (a) the pristin, (b) the V$_{\texttt{Se}}$, and (c) the Se$_{i}$, where the calculations are performed without inclusion the effect of the spin-orbit coupling (SOC). The electronic band structure of (d) the pristin, (e) the V$_{\texttt{Se}}$, and (f) the Se$_{i}$ with inclusion the effect of the SOC. The Fermi level is indicated by the dashed black lines.}
	\label{figure:Figure2}
\end{figure*}

Strong modification of the electronic properties of the PtSe$_{2}$ ML is expected to be achieved by introducing the point defect. Here, we focused on both the V$_{\texttt{Se}}$ and Se$_{i}$ because they have the lowest formation energy among the other point defects. Fig. 2 shows the calculated results of the electronic band structure of the defective systems (V$_{\texttt{Se}}$ and Se$_{i}$) compared with those of the pristin one. In the case of the V$_{\texttt{Se}}$, without inclussion of the SOC, three defect levels are generated inside the band gap [Fig. 2(b)]. Due to the absence of an anion in the V$_{\texttt{Se}}$, two excess electrons occupy the one bonding states near the valence band maximum (VBM), while the two anti-bonding states are empty, which are located close to the conduction band minimum (CBM). Our calculational results of the density of states (DOS) projected to the atoms near the V$_{\texttt{Se}}$ site confirmed that the two unoccupied antibonding states mainly originated from $d_{x{^2}+y{^2}}+d_{xy}$ orbitals of the Pt atom with a small contribution of $p_{x}+p_{y}$ orbitals of the nearest neighboring (NN) Se atoms [Fig. 3(a)]. On the other hand, admixture of the Pt-$d_{x{^2}+y{^2}}+d_{xy}$ and the small  NN Se-$p_{z}$ orbitals characterizes the one occupied bonding state. It is pointed out here that the $p$ orbitals of the next nearest neighboring (NNN) Se atoms contribute very litle to the defect states, indicating that only the Pt-$d$ and NN Se-$p$ orbitals play an important role in the defect states of the V$_{\texttt{Se}}$.  

\begin{figure}
	\centering
		\includegraphics[width=0.5\textwidth]{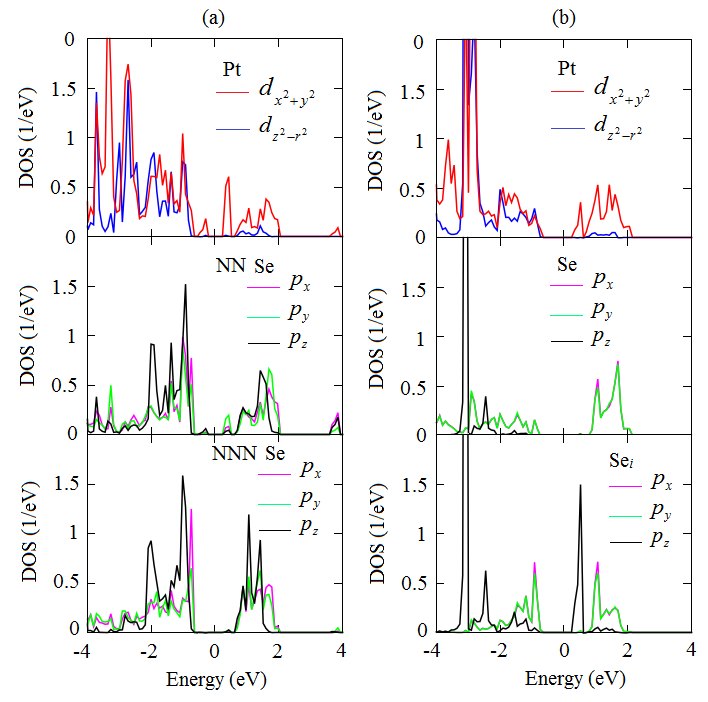}
	\caption{Density of states projected to the atomic orbitals for: (a) the V$_{\texttt{Se}}$ and (b) Se$_{i}$. The NN and NNN Se denote the nearest and next nearest neighboring Se atoms, respectively.}
	\label{figure:Figure3}
\end{figure}

The formation of the Se$_{i}$ also induces three defect levels inside the band gap, which are two occupied bonding states near the VBM and one unoccupied antibonding state close to the CBM [Fig. 2(c)]. Due to the fact that the Se$_{i}$ is on top configuration [Fig. 1(c)], the Se$_{i}$ atom forms a Se$_{i}$-Se bond with a host Se atom. In this case, the host Se atom is an anion which is in the Se$^{2-}$ oxidation state in the PtSe$_{2}$ ML, while the Se$_{i}$ is in the neutral state. Accordingly, four $p$ electron occupy the two bonding states ($p_{x}, p_{y}$) near the VBM, while the one antibonding state ($p_{z}$) located near the CBM remains empty [Fig. 3(b)]. The fully occupied two bonding level ($p_{x}, p_{y}$) and one empty antibonding level ($p_{z}$) play an important role in the nature of the Se$_{i}$-Se linear diatomic chemical bonding, which is similar to those previously reported on the sulfur interstitial of the MoS$_{2}$ ML \cite{Noh,Komsa2}.    

Turning the SOC, the energy bands are expected to develope a spin splitting, which is dictated by the lack of the inversion symmetry \cite {Rashba,Dresselhauss}. However, in the pristin PtSe$_{2}$ ML, the presence of the inversion symmetry suppresses the electronic band structures to exhibit spin polarized bands without character of the spin splitting [Fig. 2(d)]. This is supported by the fact that the absence of the spin splitting in the pristin PtSe$_{2}$ ML has been reported by Yao. $et$. $al$., using spin-angle resolved photoemission spectroscopy (Spin-ARPES) \cite {Yao}. On the other hand, introducing the Se$_{i}$ leads to the fact that the crystal symmetry of the pristin PtSe$_{2}$ ML remains unchanged [Fig. 1(d)], thus there is no spin splitting induced on the defect states [Fig. 2(f)]. 

In contrast to the Se$_{i}$ case, large spin splitting is established in the defect states of the V$_{\texttt{Se}}$ because the inversion symmetry of the pristin PtSe$_{2}$ is already broken by the stable formation of the V$_{Se}$ [Fig. 2(f)]. Figs. 4(c)-(d) show the $k$ dependence of the spin splitting corresponding to the orbital resolved of the defect states along the first Brillouin zone [Fig. 4(b)]. In the unoccupied antibonding states, the large spin splitting is observed along the $\Gamma-K$ direction, and becomes maximum at the $K$ point. On the other hand, the substantially small spin splitting is visible along the $\Gamma-M$ direction [Fig. 4(c)]. Conversly, a complicated trend of the spin splitting is observed in the occupied bonding state: the spin splitting is small at the $K$ point, and rises continueously up to maximum at midway between the $K$ and $\Gamma$ points, but gradually decreases until the zero spin splitting is achieved at the $\Gamma$ point. The similar trend of the spin splitting is also visible along the $\Gamma-M$ direction. It is noted here that zero spin splitting is observed at the $\Gamma$ and $M$ points due to time reversability. We  identify the spin splitting on the defect states at the $K$ point: $\Delta_{K(1)}=152$ meV and $\Delta_{K(2)}=127$ meV in the lower and upper unoccupied antibonding states, respectively, and $\Delta_{K(3)}=5$ meV in the occupied bonding states. The large spin splitting found in the unoccupied antibonding states ($\Delta_{K(1)}$ and $\Delta_{K(2)}$) is comparable with those found in the pristin MoS$_{2}$ ML (148 meV \cite {Zhu}) and defective WS$_{2}$ ML (194 meV \cite {Li}), but is much larger than that of convensional semiconductor III-V and II-VI quantum well ($<30$ meV \cite {Gui,Nitta}). Indeed, they are fully comparable to the recently reported surface Rashba splitting (of some 100 meV) observed on Au (111) \cite {LaShell}, Bi (111) \cite {Koroteev}, PbGe (111) \cite {Yaji}, Bi$_{2}$Se$_{3}$ [001] \cite {King}, and W [110] \cite {Hochstrasser} surfaces. 

\begin{figure}
	\centering
		\includegraphics[width=0.5\textwidth]{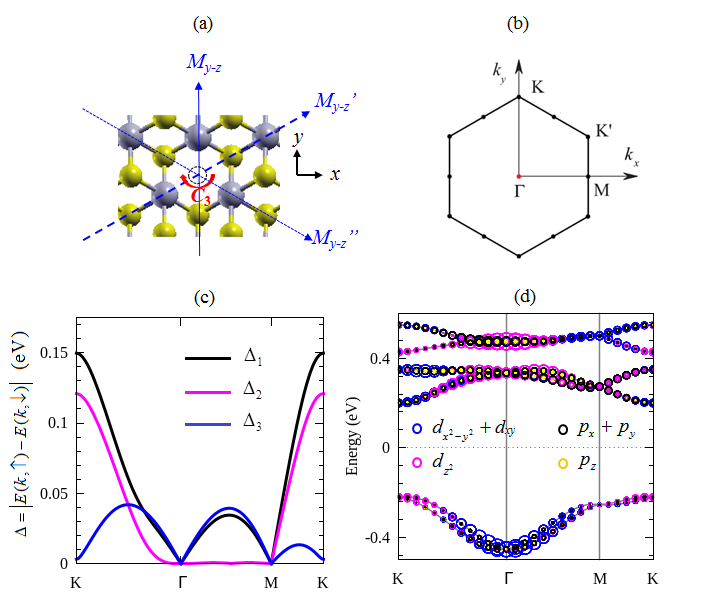}
	\caption{Relation between the point group symmetry of the V$_{\texttt{Se}}$, the spin splitting, and the orbital contributions to the defect states in the first Brillouin zone. (a) Symmetry operation in the real space of the V$_{\texttt{Se}}$ corresponding to (b) the first Brillouin zone. (c) The spin splitting in the defect states calculated along the first Brillouin zone. Here, the $\Delta_{1}$ and $\Delta_{2}$ represent the spin splitting for the lower and upper unoccupied antibonding states, respectively, while $\Delta_{3}$ represents the spin splitting of the occupied bonding state. (d) Orbital-resolved of the electronic band structures calculated in the defect states. The radius of circles reflects the magnitudes of spectral weight of the particular orbitals to the band.}
	\label{figure:Figure4}
\end{figure}

To clarify the origin of the observed spin splitting, we consider orbitals contribution to the defect states of the V$_{\texttt{Se}}$ projected to the bands strcutures in the first Brillouin zone as shown in Fig. 4(d). We find that the unoccupied antibonding states reveal strong hybridization between the Pt-$d_{x{^2}+y{^2}}+d_{xy}$ and Se-$p_{x}+p_{y}$ orbitals at the $K$ point, which induces the large spin splitting at the $K$ point. Toward the $\Gamma$ point, these contributions are gradually replaced by the hybridization between Pt-$d_{z^{2}}$ and Se-$p_{z}$ orbitals, which contributes only minimally to the spin splitting around the $\Gamma$ point. The same orbitals hybridization are also visible in the $K$ point of the occupied bonding state, resulting in that the spin splitting becomes very small. However, around midway between the $K$ and $\Gamma$ points, the contribution of the Pt-$d_{x{^2}+y{^2}}+d_{xy}$ and Se-$p_{x}+p_{y}$ orbitals to the occupied bonding state increases, which enhances the spin splitting around the $\Gamma$ point. Remarkably, the in-plane orbital hybridizations [Pt-$d_{x{^2}+y{^2}}+d_{xy}$ and Se-$p_{x}+p_{y}$ orbitals] in the defect states induced by the V$_{\texttt{Se}}$ play an important role for inducing the large spin splitting. 

To further reveal the nature of the spin splitting in the defect states of the V$_{\texttt{Se}}$, we consider our system based on the symmetry arguments. As mentioned before that the structural relaxation retains the symmetry of the V$_{\texttt{Se}}$ becomes the  $C_{3v}$ [Figs. 1(b) and 4(a)]. Here, the symmetry itself consists of a $C_3$ rotation and a mirror symmetry operation $M_{y-z}:x\longrightarrow-x$, where $x$ is along the $\Gamma$-$M$ direction [Fig. 4(b)]. Therefore, the spin splitting for general $k$ is determined by time reversal symmetry and the  $C_{3v}$ point group symmetry. By using the theory of invariants, the  $C_{3v}$ leads to the spin splitting \cite {Zhu,Fu,Absor4}:
\begin{equation}
\label{4}
\Delta (k,\theta)=\left(\alpha^{2}(k) + \beta^{2} (k)\sin^{2}(3\theta)\right)^{1/2}
\end{equation} 
Here, $\alpha (k)$ and $\beta (k)$ are the coefficient representing the contribution of the in-plane and out-of-plane potential gradient asymmetries, respectively, and $\theta =\tan^{-1}(k_{y}/k_{x})$ is the azimuth angle of momentum $k$ with respect to the $x$ axis along the $\Gamma$-$K$ direction. In Eq. (\ref{4}), due to the $\left|\sin (3\theta)\right|$ dependent of the $\Delta$, the spin splitting is minimum when $\theta=n\pi/3$, where $n$ is an integer number. Therefore, it is expected that the small spin splitting is observed along the $\Gamma$-$M$ direction. On the other hand, the spin splitting becomes maximum when $\theta=(2n+1)\pi/6$, which can be visible along the $\Gamma$-$K$ direction. These predicted spin splittings along the $\Gamma$-$M$ and the $\Gamma$-$K$ directions are in fact consistent with our calculational results of the spin splitting in the defect states shown in Fig. 4(c).

Thus far, we found that the spin-orbit splitting in the electronic band structures of the PtSe$_{2}$ ML can be induced by introducing the point defects. Considering the fact that the large spin splitting is achieved on the $K$ point of the unoccupied antibonding states, $n$-type defective PtSe$_{2}$ ML for spintronics is expected to be realized. This is supported by the fact that a deep single acceptor induced by the chalcogen vacancy has been predicted on MoS$_{2}$ ML \cite {Noh}. Moreover, the observed large splittings enable us to allow operation as a spintronics device at room temperature \cite {Yaji,King}. As such our finding of large spin splitting are useful for realizing spintronics application of the PtSe$_{2}$ ML system. 

It is pointed out here that our proposed approach for inducing the large spin splitting by using the point defects is not only limited on the PtSe$_{2}$ ML, but also can be extendable to other $T-MX_{2}$ ML systems such as the other platinum dichalcogenides like PtS$_{2}$ and PtTe$_{2}$ \cite {Manchanda}, vanadium dichalcogenide like VSe$_{2}$, VS$_{2}$, VTe$_{2}$ \cite {Cudazzo}, and rhenium disulfides (ReS$_{2}$) \cite {Horzum}, where the electronic structure properties are similar. Importantly, controlling the electronic properties of these materials by using the point defects has been recently reported\cite {Manchanda, Horzum}. Therefore, this work paves a possible way to engineer the spin splitting properties of the two-dimensinal nanomaterials, which provide useful information for the potential applications in spintronics.

\section{CONCLUSION}

We have investigated the spin-orbit-induced spin splitting in the defective of the PtSe$_{2}$ ML systems by employing the first-principles density functional theory (DFT) calculations. First, we have obtained the formation energy of the native point defects and found that both the Se vacancy (V$_{\texttt{Se}}$) and Se interstitial (Se$_{i}$) are the most stable defects formed in the PtSe$_{2}$ ML. By taking into account the effect of the spin-orbit coupling in our DFT calculations, we have found that the large spin-orbit splitting (up to 152 meV) is observed in the defect states induced by the V$_{\texttt{Se}}$. We have clarified the origin of the spin splitting by considering orbital contributions to the defect states, and found that the large spin splitting is induced by strong hybridization between the Pt-$d_{x{^2}+y{^2}}+d_{xy}$ and Se-$p_{x}+p_{y}$ orbitals. Recently, the defective of the PtSe$_2$ ML has been extensively studied \cite {WZhang,Komsa1,Zulfiqar}. Our study clarifies that the defects play an important role in the spin splitting properties of the PtSe$_2$ ML, which is important for designing future spintronic devices.

\begin{acknowledgments}

This work was partly supported by the Fundamental Reserach Grant (No. 2237/UN1.P.III-DITLIT-LT/2017) funded by the ministry of research and technology and higher eduacation, Republic of Indonesia. Part of this reserach was supported by BOPTN reserach grant (2017) founded by Faculty of Mathematics and Natural Sciences, Universitas Gadjah Mada. The computations in this research were performed using the high performance computing facilities (DSDI) at Universitas Gadjah Mada, Indonesia. 

\end{acknowledgments}

\bibliography{Reference1}


\end{document}